\documentclass[cameraready]{Interspeech}
\newcommand{\plainfootnote}[1]{
  \insert\footins{
    \hsize=\columnwidth
    \parindent=1em
    \small #1\par
  }
}

\title{The silence of the weights: a structural pruning strategy for attention-based audio signal architectures with second order metrics}

\author[affiliation={1}, orcid=0009-0000-2153-2485, correspondingauthor]{Andrea}{Diecidue}
\author[affiliation={2}, orcid=0000-0001-9512-0440]{Carlo Alberto}{Barbano}
\author[affiliation={1}]{Piero}{Fraternali}
\author[affiliation={3}, orcid=0000-0002-7657-6271]{Mathieu}{Fontaine}
\author[affiliation={3}, orcid=0000-0003-4274-8298, correspondingauthor]{Enzo}{Tartaglione}

\address{
    $^1$ Politecnico di Milano, Italy \\
    $^2$ University of Turin, Italy \\
    $^3$ LTCI, Télécom Paris, Institut Polytechnique de Paris, France
}

\email{andrea.diecidue@polimi.it, enzo.tartaglione@telecom-paris.fr}


\usepackage{comment,amsmath,graphicx,hyperref, amssymb, multirow, booktabs, makecell, comment, algorithm, algpseudocode, hyperref, subfigure}

\begin{document}

\maketitle

\begin{abstract}
Transformer-based models have become the state of the art across multiple domains, from natural language processing to machine listening, thanks to the attention mechanisms. However, the attention layers require a large number of parameters and high-end hardware for both training and inference. We propose a novel channel-pruning technique explicitly targeted at the attention mechanism, decoupling the pruning of each head and the four layers in the attention block: query, key, value, and output projection matrices, employing a second-order metric to score the network's parameters. We compare our technique against head-pruning strategies and magnitude-driven scoring metrics, investigating the effects of pruning on  Audio Spectrogram Transformer (AST) and Whisper. Our results show that even after pruning 50\% of the parameters in the attention block, performance is largely preserved. The code will be made available upon 
acceptance of the paper.
\plainfootnote{Paper under review.}
\end{abstract}

\section{Introduction}
\label{sec:intro}

Transformer-based models~\cite{attention} have set the state of the art across a wide range of domains, including computer vision~\cite{vit}, natural language processing~\cite{bert}, and more recently, machine listening~\cite{Qwen-Audio}.
In the audio field, transformer-based architectures are applied to tasks such as sound classification~\cite{ast,audiomae}  as well as source separation, with models like TF-GridNet~\cite{tfgridnet} and TF-Locoformer~\cite{tflocoformer}, and machine transcription or translation like Whisper~\cite{whisper}. 

Transformers are known to scale effectively with model and dataset size~\cite{scalinglaw}. This led to the development of large architectures with up to a billion parameters~\cite{stableaudioopen, gpt, llama}.
However, the high number of parameters leads to longer training times, as well as increased memory and energy requirements for storage and execution, which limits the devices that can support the network.
To address this challenge, pruning techniques were designed to remove redundant parameters without affecting the network performance.

Pruning~\cite{pruning_survey} is based on the idea that deep neural networks (DNNs) are over-parametrised with respect to downstream tasks.
There are two main approaches to pruning: structured and unstructured. In structured pruning, entire neurons, convolutional filters, attention heads~\cite{sixteenheadsbetterthanone, specializedheads}, or even layers, are removed. Unstructured pruning, on the other hand, focuses on each parameter independently.
Although structured pruning is less flexible when it comes to which parameters to prune, it can directly speed up the network forward pass by changing the topology of the network, without the need for special hardware like unstructured pruning does~\cite{DeepCompression}. 
In the case of transformers, a third pruning technique, called token pruning~\cite{dynamicvit, tome, fastv},
consists of removing or merging tokens that become redundant at specific points in the network, thus reducing the number of inputs in the following layers. 

In machine listening, various strategies have been explored. PARP~\cite{Lai2021PARP} applies unstructured weight pruning to speech models but produces irregular sparsities that do not lead to any real speed-up.
Consequently, structured pruning has gained attention: for instance, Peng et al.~\cite{Peng2023Structured} applied L0-based structured pruning to Wav2Vec~2.0, eliminating whole attention heads and feed-forward channels and achieving $\approx 50\%$ compute reduction with negligible performance loss. Other works, such as Wang et al.~\cite{Wang2023TaskAgnostic}, similarly introduced fine-grained attention head pruning to mitigate accuracy loss in structured compression, achieving a 72\% parameter reduction and a $2\times$ speedup with no significant performance drop. Additionally, token pruning has been extended to audio Transformers: dropping low-importance spectrogram tokens can reduce attention computations by 30–40\% with an under 1\% accuracy loss~\cite{Lee2025TokenPruning}. Dynamic pruning methods further adapt the model at inference: \cite{Someki2025ContextAware}, for example, proposes context-driven gating that skips less critical attention blocks on-the-fly, preserving performance while reducing latency by $\approx 30\%$.
However, structured pruning of attention modules in audio remains underexplored. Most prior works remove entire heads or tokens; few address finer-grained pruning (e.g., Q/K/V channels), a direction explored in NLP Transformers~\cite{kvpruner} but underexplored in audio.

In this paper, we propose a channel-wise structural pruning strategy specifically targeting the attention block, and the use of Fisher information as a scoring metric for parameters. We compare against more standard head-wise pruning and magnitude-based scoring methods. We present the results of our method on two tasks: audio classification using the AST~\cite{ast} architecture and audio transcription/translation using the Whisper architecture.
Our results show that it is possible to reduce the number of parameters in the attention layers by 50\% with only a very minor loss in performance.

\section{Proposed method}
\label{sec:method}
\begin{figure}[t]
    \centering
    \includegraphics[width=\linewidth]{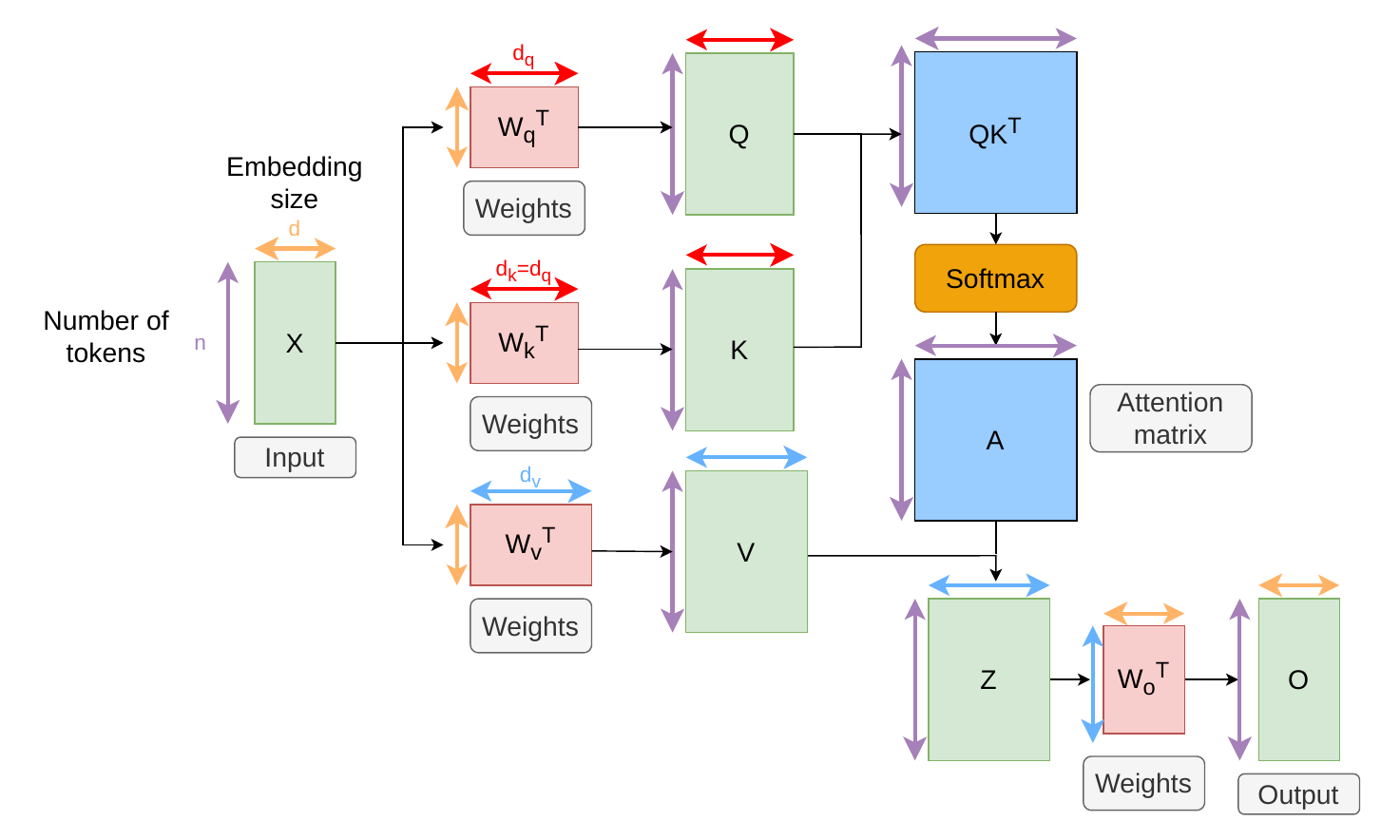}
    \caption{Notation for the self-attention block.}
    \label{fig:self-attn}
\end{figure}

\paragraph*{Self-attention block description}
The attention block, as in Fig.~\ref{fig:self-attn}, is composed of four weight matrices: $\mathbf{W_q} \in \mathbb{R}^{d_q \times d}$, $\mathbf{W_k} \in \mathbb{R}^{d_q \times d}$, $\mathbf{W_v} \in \mathbb{R}^{d_v \times d}$, $\mathbf{W_o} \in \mathbb{R}^{d \times d_v}$ representing the queries, keys, values, and output projection weights. 

\begin{figure}[t]
    \centering
    \includegraphics[width=\linewidth]{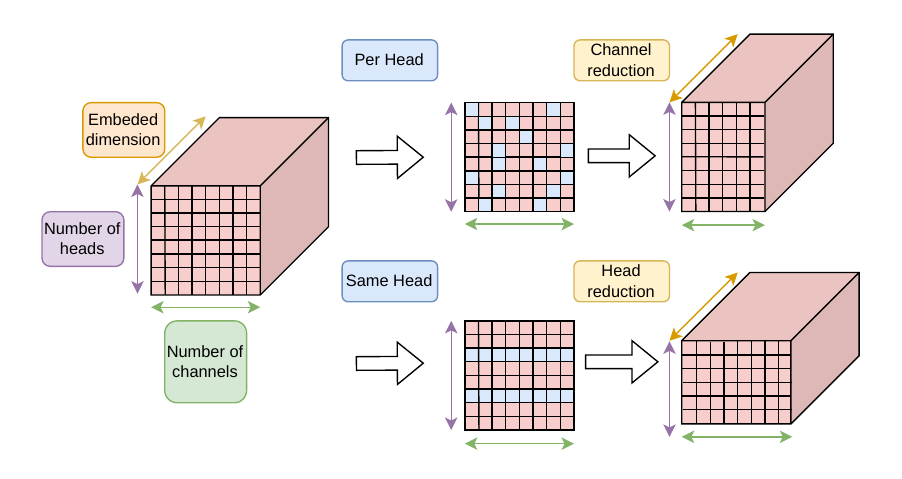}
    \caption{Schemes of per-head and same-head pruning patterns, in order.}
    \label{fig:pruning-patterns}
\end{figure}

Although from the implementation perspective these are usually treated as 2D matrices, there are actually three parameters that control their shape: the embedded dimension $\mathbf{d}$, the number of heads $\mathbf{n_h}$, and the number of channels $\mathbf{n_c}$, as in Fig.~\ref{fig:pruning-patterns}. Most attention implementations use $\mathbf{n_c} = \mathbf{d}/\mathbf{n_h}$, making the matrix squared and thus more computationally efficient. In the case of self-attention, usually $\mathbf{d_q}= \mathbf{d_v} =\mathbf{n_h} \cdot \mathbf{n_c}$ to make also the entire operation more efficient.

Motivated by~\cite{schubert}, the following observations and hypotheses are assumed.
First, not all layers require the same number of parameters. In fact, the number of heads and channels can vary across layers. Second, the projection matrices $\mathbf{W_q}, \mathbf{W_k}, \mathbf{W_v}, \mathbf{W_O}$ do not need to be pruned in the same way.
In fact, only $\mathbf{Q}$ and $\mathbf{K}$ must have one equal-length side to perform the dot product. 
In the same way, $\mathbf{W_o}$ in Fig.~\ref{fig:self-attn} processes $\mathbf{Z} \in \mathbb{R}^{n \times d_v}$ to produce the output of $\mathbf{O} \in \mathbb{R}^{n \times d}$, so they share the same dimension $\mathbf{d_v}$.
This implies that, when pruning the parameters in the layer, we only have two constraints: $\mathbf{W_q}$ and $\mathbf{W_k}$ must have the same output dimension, and $\mathbf{W_v}$'s output dimension must match $\mathbf{W_o}$'s input dimension.
While each head may have a different number of channels, keeping the number of channels constant across the heads of a single layer is computationally more efficient. 

\paragraph*{Pruning scheme}
Pruning along the head dimension (\textbf{EH}) is the preferred method in the literature~\cite{ sixteenheadsbetterthanone, specializedheads, Wang2023TaskAgnostic} for two main reasons: first, most attention implementations are optimized for the head dimension, so removing entire heads yields greater benefits in terms of inference speed. Second, heads are independent subspaces, so pruning entire heads usually hinders performance less. Nonetheless, channel-wise pruning can be beneficial; it still reduces inference speed while achieving competitive performance. 
In this direction, we propose a novel pruning scheme that independently selects, for each head, the set of channels to prune, to preserve the most important dimension of each head. Fig.~\ref{fig:pruning-patterns} shows the difference in pruning patterns between per-head pruning and head-wise pruning.
We formulate per-head channel pruning (\textbf{PH}) as a budget allocation problem, aiming to select the number of channels to achieve the desired sparsity while minimising the score of the pruned parameters. To solve the problem, we employ a greedy approach that, for each layer, defines a total budget (the number of channels to remove in QK and VO), computes the importance score for each channel, and iteratively selects the channels to prune. At each step, the set of channels with the lowest cost that fits within the budget is selected, and the set of possible channels for that layer is updated.

\paragraph*{Scoring parameter metric}
In addition to our pruning scheme, we also adopt a second-order metric to score the parameters. Most structural pruning approaches use the raw parameter magnitudes (\textbf{MAG}), typically computed as norms (L1 or L2), to rank them. The parameters with the lowest scores are pruned because they are deemed less impactful on the network's output. On the other hand, parameter magnitudes can vary substantially across layers; layers closer to the input tend to have smaller magnitudes than those in later layers, which favors pruning earlier layers. Fisher information (\textbf{FI}) is very attractive as a metric because not only it is less sensitive to the scaling problem, but it also minimizes total the loss of information in the network.
Fisher information measures the amount of information each parameter in the network carries about the output. The lower the value is, the less impactful the parameter is. We can show this starting from the formulation of Fisher Information as:

\begin{equation}
\mathcal{I}(\theta) = - \mathbb{E} \Bigg[ \frac{\partial^2 \log f(X;\theta)}{\partial \theta^2} | \theta \Bigg].
\end{equation}
where $\theta$ is an unknown parameter, $X$ is an observable random variable, and $f(X;\theta)$ is a probability distribution function. 
Fisher information measures the variance of the score of $\theta$ with respect to $X$. Low variance indicates low information. To estimate the importance of a single parameter of a neural network, we use a loss function that represents the log-likelihood in place of $logf(X;\theta)$, where $X$ is a set of samples, and $\theta$ is the parameter.
Thus, Fisher information of $\theta$ can be computed by feeding a set of samples $X$ to a neural network as:

\begin{equation}
\mathcal{I}(\theta) = - \mathbb{E} \Bigg[ \left( \frac{\partial \mathcal{L}(\theta)}{\partial \theta} \right)^2 \Bigg].
\end{equation}

Depending on the chosen parameter grouping, we can accumulate per-parameter values to estimate the Fisher information carried by the entire parameter group under examination.
Other works have explored second-order information, such as the full Hessian matrix, but Fisher encodes nearly the same information at a much lower computational cost. Computing the full Hessian matrix is quadratic in the number of samples, while Fisher's is only linear.

\paragraph*{Thresholding}
Finally, we will explore two threshold strategies: global (\textbf{G}), which consolidates all channels or heads across layers and decides which to prune based on their scores, and local (\textbf{L}), which treats each layer independently.
With the first strategy, we can prune each layer by different amounts. In contrast, in the second case, each layer will be pruned by the same amount because all attention blocks at the beginning have the same number of heads, channels, and embedded dimension.

\section{Results}
\label{sec:results}
\paragraph*{Experimental setup}
We test our pruning approach on two architectures: AST for audio classification tasks and Whisper for machine transcription and translation. For AST, we present results on Audioset (the balanced version available on \href{https://huggingface.co/datasets/agkphysics/AudioSet}{Huggingface}) and SpeechCommand 2.
For Whisper, we use the "medium" model, with weights provided by OpenAI via \href{https://huggingface.co/openai/whisper-medium}{Hugging Face} and evaluate on both transcription and translation tasks.
We test all combinations of pruning schemes, scoring metrics, and threshold approaches on both architectures. Pruning is applied iteratively for $10$ steps; at each iteration, we prune 10\% of the parameters from the attention blocks. 
For AST, we fine-tune the model after each iteration using LoRA~\cite{lora} for $3$ epochs with a learning rate of $10^{-4}$ and AdamW optimizer. 
For Whisper, we use a mix of LibriSpeech \cite{librispeech}, MultiLingual LibriSpeech (MLS) \cite{MLS}, CommonVoice \cite{commonvoice}, VoxPopuli \cite{voxpopuli}, FLEURS \cite{fleurs}, and CoVoST \cite{covost}, totalling 33k hours of audio for fine-tuning, with a learning rate of $10^{-4}$ and SGD optimizer.
\begin{figure}
    \centering
    \includegraphics[width=.45\textwidth]{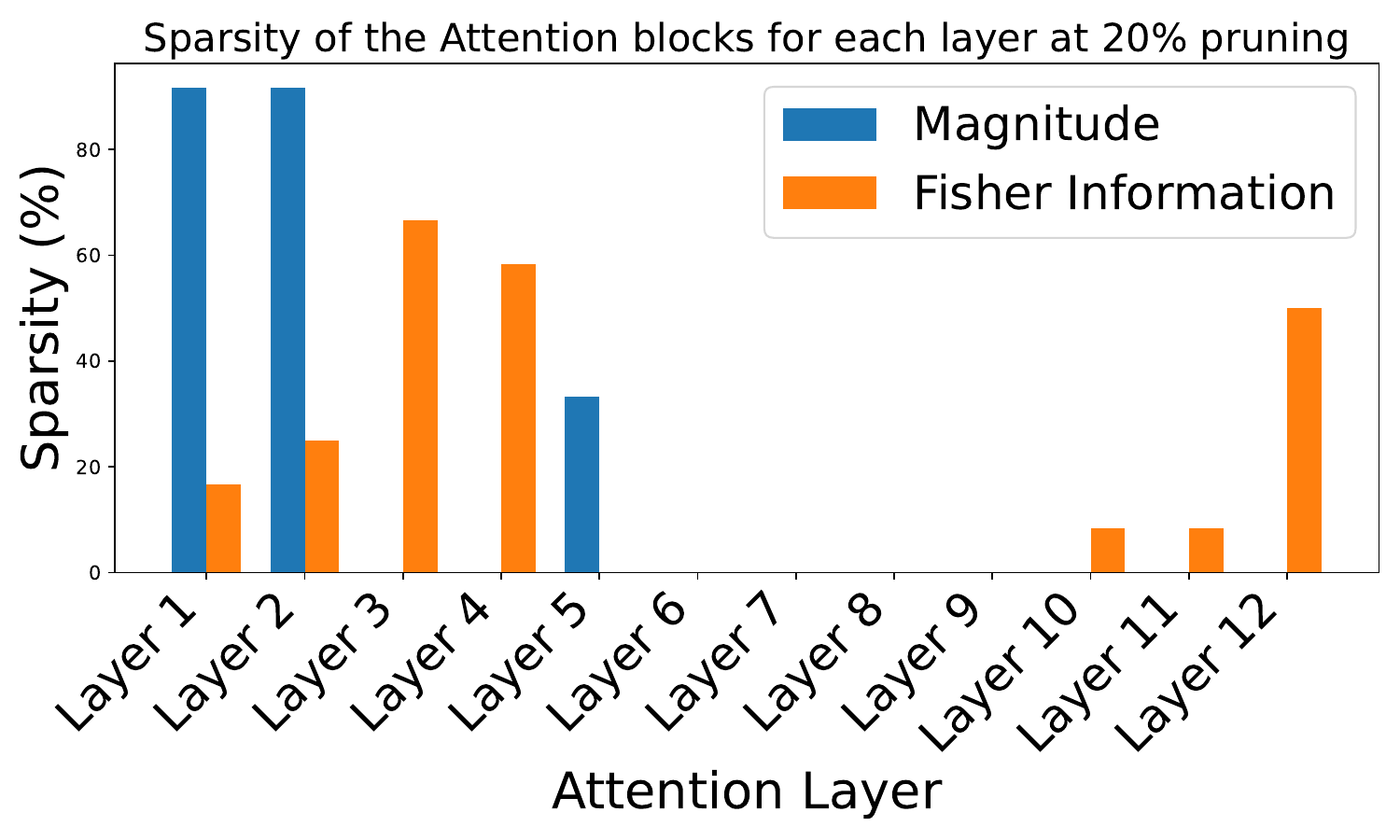}
    \caption{Sparsity in the attention blocks of each layer for the per-head pruning scheme approach on the SpeechCommands dataset. Layers are ordered from the input to the output of the network. Magnitude tends to favor earlier layers, whereas Fisher distributes the budget differently, yielding better models.}
    \label{fig:sparsity_layers}
\end{figure}
\begin{figure}
    \centering
    \includegraphics[width=.45\textwidth]{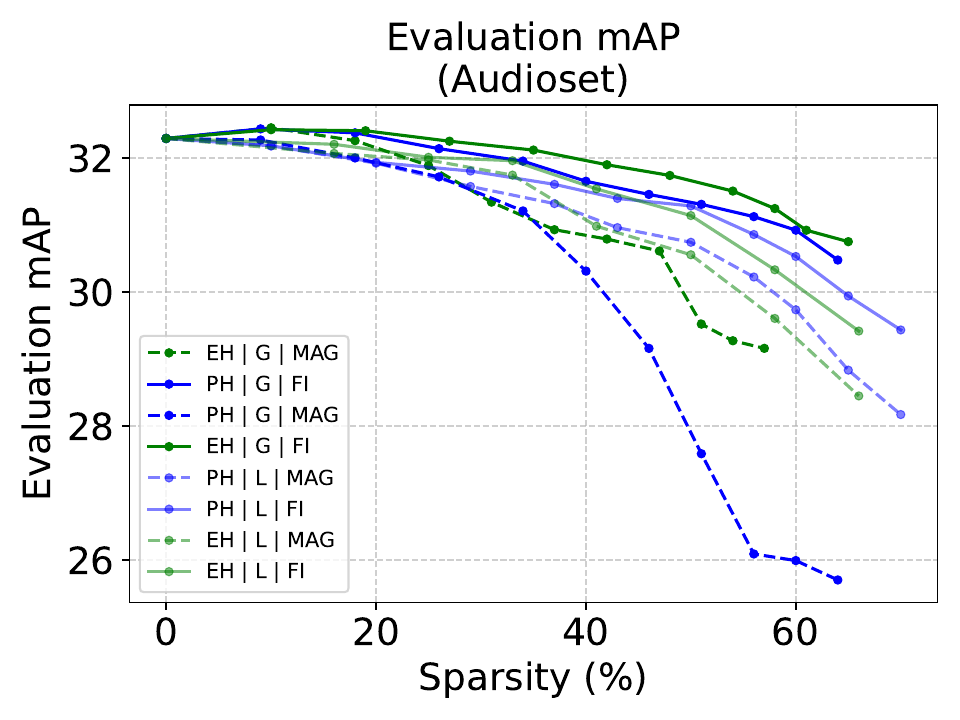}
    \caption{mAP of AST on Audioset at different sparsity rates.}
    \label{fig:eval_audioset}
\end{figure}
\begin{figure}
    \centering
    \includegraphics[width=.45\textwidth]{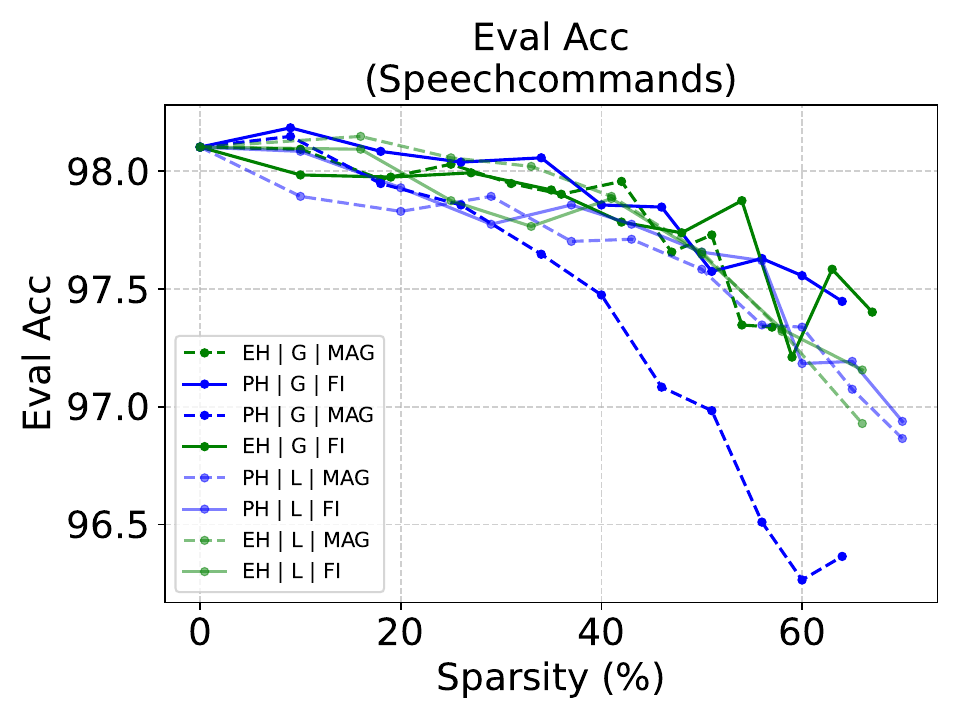}
    \caption{Accuracy of AST on Speechcommands at different sparsity rates.}
    \label{fig:eval_speechcommands}
\end{figure}
\begin{figure}
    \centering
    \includegraphics[width=.45\textwidth]{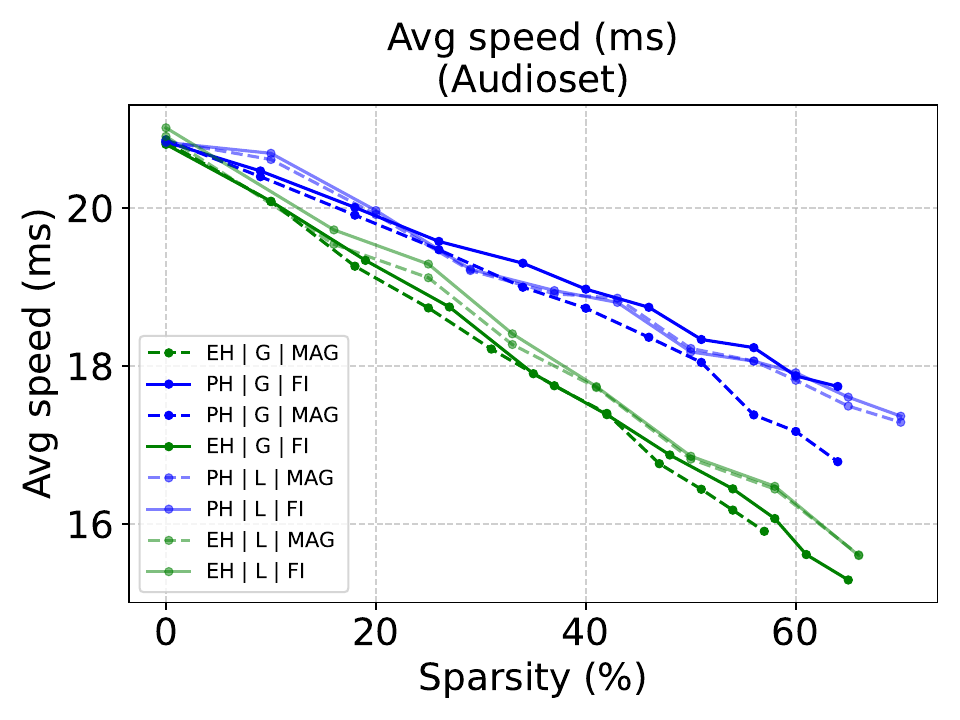}
    \caption{Inference speed on a single sample from the Audioset dataset at different pruning levels, averaged over 100 samples.}
    \label{fig:speedup}
\end{figure}
\paragraph*{Importance score}
Fig.~\ref{fig:sparsity_layers} shows how Fisher information and magnitude select which layers to prune. Magnitude favors layers closer to the input because they have a smaller scale than later layers. Fisher, on the other hand, selects both the later and earlier layers. We believe that pruning in earlier layers occurs because MEL spectrograms are very sparse images, making the features learned by those layers more easily condensed.
\paragraph*{Main pruning results}
Fig. \ref{fig:eval_audioset} and \ref{fig:eval_speechcommands} report the results of the different pruning techniques applied to the AST architecture on Audioset and Speechcommands, respectively. As expected, when we use Fisher information to score the parameters, the model performs better than when we prune by magnitude. We also observe that the proposed method outperforms all magnitude-based methods and is on par with head-wise pruning using Fisher information. 
Unsurprisingly, magnitude-based pruning works better with local thresholding than with global thresholding, as local thresholding sidesteps the scale difference across layers. Conversely, Fisher information is not subject to the scale problem and performs better with global thresholding. When it comes to inference speed (Fig.~\ref{fig:speedup}), we notice that head-wise pruning is 1-2ms faster with respect to per-head pruning. This is not surprising; head-wise pruning removes entire scaled dot product operations from the computation graph, while per-head pruning only reduces the computations for each scaled dot product. 

\begin{figure}
    \centering
    \includegraphics[width=.45\textwidth]{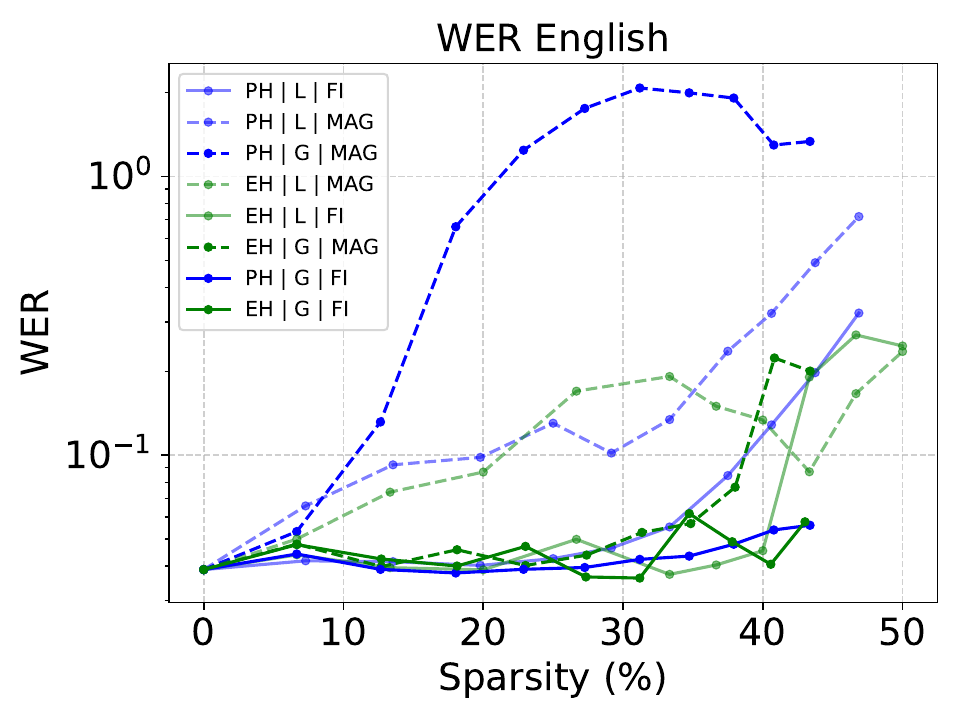}
    \caption{WER on Librispeech English dataset at different pruning iterations, with logarithmic scale (lower is better).}
    \label{fig:wer_en}
\end{figure}

\begin{figure} 
    \centering 
    \includegraphics[width=.45\textwidth]{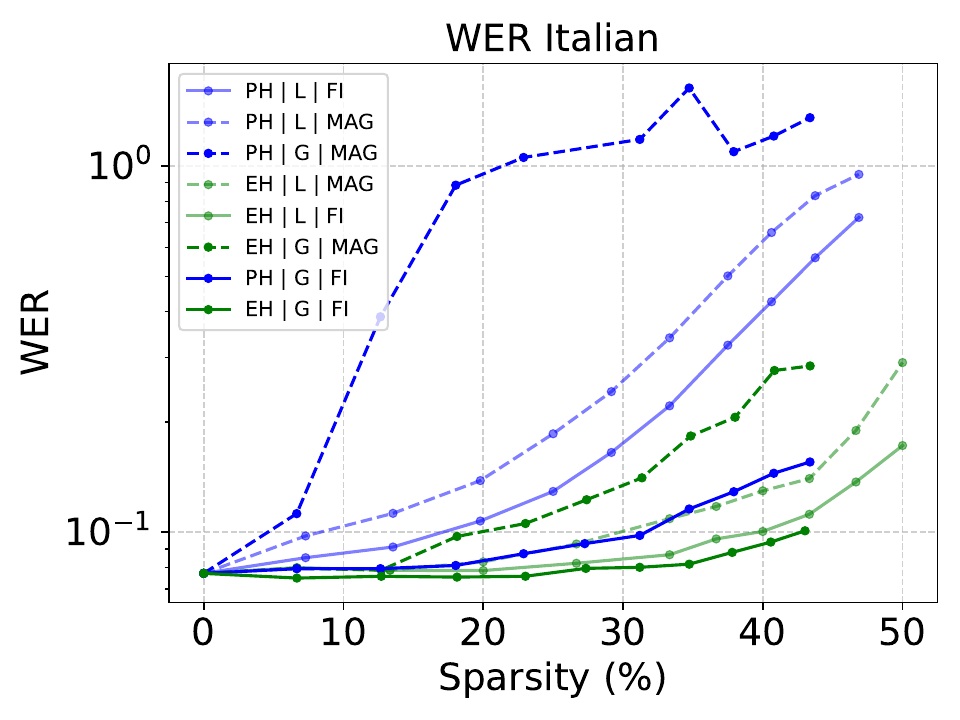} 
    \caption{WER on CommonVoice Italian dataset at different pruning iterations, with logarithmic scale (lower is better).} \label{fig:wer_it} 
\end{figure}

\begin{figure}[t] 
    \centering 
    \includegraphics[width=.45\textwidth]{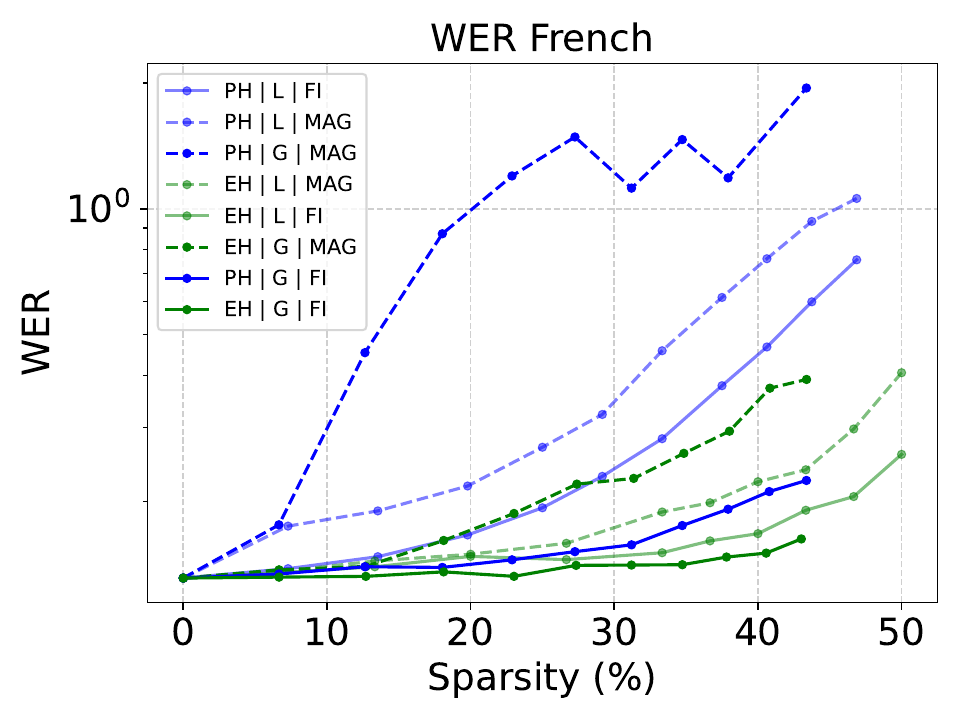} 
    \caption{WER on CommonVoice French dataset at different pruning iterations, with logarithmic scale (lower is better).}
    \label{fig:wer_fr} 
\end{figure}

\begin{figure}[t] 
    \centering 
    \includegraphics[width=.45\textwidth]{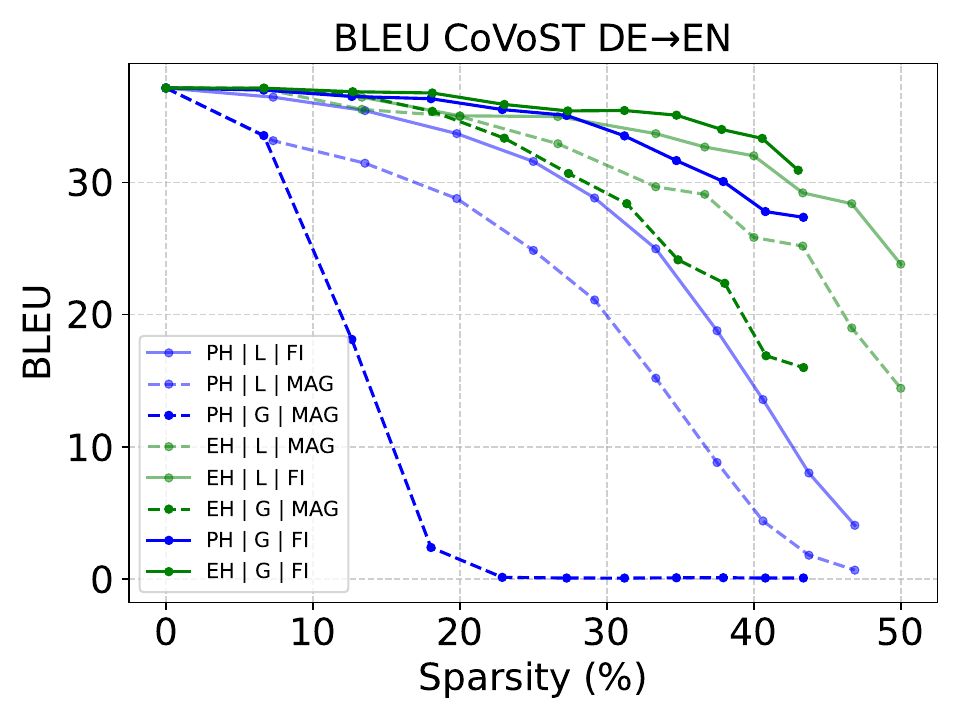} 
    \caption{BLEU on CoVoST German to English dataset at different pruning iterations (higher is better).} 
    \label{fig:blue_de_en} 
\end{figure}

Similar observations can be done for Whisper. Fig.~\ref{fig:wer_en},~\ref{fig:wer_it}, and~\ref{fig:wer_fr} report the word error rate (WER) on Librispeech English and on the Italian and French subsets of CommonVoice. Our per-head pruning scheme is competitive with head-wise pruning in terms of performance, and is within 1\% of the original model. When it comes to machine translation (Fig.~\ref{fig:blue_de_en}), we observe a larger drop in performance. It is worth noting that although we train the model on these languages, the training set is fairly small (about 1500 samples per language for transcription and 1000 for translation), so we attribute the performance primarily to the scoring metric, which retains the most important parameters. 

    
    
    
    

\section{Conclusions}
\label{sec:conclusions}
In this paper, we present a novel pruning pattern for the channel dimension in the Attention block, combined with Fisher Information to rank the parameters. In particular, we evaluate the effect of this type of pruning on AST and Whisper architectures, comparing it against more standard techniques, such as head-wise pruning and magnitude ranking. From the experimental results, it appears that careful channel-wise pruning can be on par with head-wise pruning, especially when combined with Fisher Information.

Several future directions arise from this work, including exploring other fields, such as CV and NLP, and pruning simultaneously along both the head and channel dimensions. In particular, we believe that relaxing the constraint of keeping the same number of channels across all heads could yield a more optimal pruning pattern, although careful optimization of the scaled dot product is needed to avoid bottlenecks that would hinder inference speed.
\newpage
\section{Acknowledgments} 
This work was supported by the French National Research Agency (ANR) in the framework of the JCJC project “BANERA” under Grant ANR-24-CE23-4369, from the European Union’s Horizon Europe Research and Innovation Programme under grant agreement No. 101120237 (ELIAS), and by the Hi! PARIS Center on Data Analytics and Artificial Intelligence.

We acknowledge VSB – Technical University of Ostrava, IT4Innovations National Supercomputing Center, Czech Republic, for awarding this project access to the LUMI supercomputer, owned by the EuroHPC Joint Undertaking, hosted by CSC (Finland) and the LUMI consortium through the Ministry of Education, Youth and Sports of the Czech Republic through the e-INFRA CZ (grant ID: 90254).

\bibliographystyle{IEEEtran}
\bibliography{mybib}
\section{Generative AI Use Disclosure}
Generative AI was used only to refine the paper from a syntactic and semantic perspective; it was not used to generate any plots, references, or other significant parts of the paper. 
\end{document}